# Kara-Kichwa Data Sovereignty Framework

## *Reference Point for Indigenous Data Authority Renaissances in LAC*

WariNkwi K. Flores[1-2], KunTikzi Flores[3], Rosa M. Panama[4], KayaKanti Alta[1-3]

[1]Kinray Hub: Indigenous RDI Think-Do Tank, Cotacachi, Imbabura, Ecuador

[2]The University of Arizona: School of Natural Resources and the Environment, Tucson, AZ, United States

[3]Community of Arrayanes & Santa Barbara, Clan Chichup'mp', Kara-Kichwa Nation, Cotacachi, Imbabura, Ecuador

[4]Community of Įltakī, Kara-Kichwa Nation, Cotacachi, Imbabura, Ecuador

https://orcid.org/0009-0001-9731-6217

**Abstract**

For Indigenous Peoples of the Apŷa Yala *(or Abya Yala)*, particularly in the Kara and Kichwa citizens of the Pan-Andean–Amazonian biocultural region, data is not merely a knowledge or information resource; it is the extension of ***Khipu Panaka*** (Indigenous data authority), treading the data lifecycle – genealogical and relational memory held within customary law and collective responsibility. This perspective paper presents the Kara-Kichwa Data Sovereignty Framework, a living legal–ethical instrument developed through autopoietic Indigenous storytelling, rights to story and place, and Indigenous-informed scope review to engage with external Indigenous data frameworks, counteracting "intellectual gentrification" and the systemic invisibility of Andean-Amazonian Indigenous Peoples within global digital transformation. The framework codifies five customary pillars – ***Kamachŷ*** (self-determination: community owns data about itself), ***Ayl'u-l'aktapak kamachŷ*** (collective authority and polygovernance), ***Tantanakuy*** (collective deliberation and relational accountability), ***Wil'ay-panka-tantay*** (physical custody of data and knowledge confidentiality), and ***Sumak kawsay*** (biocultural ethics and intergenerational responsibility) – to guide the data lifecycle from generation to expiration. While this framework arises from Kara-Kichwa customary law, the pillars outline how its governance logics serve as a *reference point for Indigenous data authority renaissances in Latin America and the Caribbean (LAC)*, through respectful adaptation by other Indigenous nations on their own terms.

**Positionality Rationale**: For time immemorial, Indigenous Peoples of Apŷa Yala have developed technological innovations (i.e., *khipu* technological systems) to conceptualize, collect, store, govern, and interpret their own and territorial data [1–4]. For +533 years, the doctrine of discovery has justified the dispossession of Indigenous commonwealths and the intellectual gentrification of their commonwealths, worldviews, cultures, science, technology, and societies [5]. In LAC, the lack of Indigenous data sovereignty instruments has perpetuated the invisibility and marginalization of Indigenous Peoples' laws and authority, reinforcing historical systemic dispossession [6]. Its absence from the Sustainable Mountain Development recommendations to close the data gaps exacerbates this intellectual gentrification, turning Andean communities charity of their own knowledge [7]. Also, the WEIRD (western, educated, industrialized, rich, "democratic") science builds belief systems that LAC Communities cannot manage their data standards, thereby systematically excluding these Communities from national and international negotiations and decision-making on *Pan-Andean-Amazonian-Atlantic* governance [8,9].

The Kara-Kichwa Data Sovereignty framework is an Indigenous data authority *(Khipu Panaka)* renaissances reference point to these challenges in LAC. This authority is collective and exercised by each Community assembly, led by its elected Indigenous representatives in a consensus-driven governance institution. It aims to ground the IEEE Std 2890 in local realities – from statistics, research, development, and innovation to knowledge domains and their derived data, such as digital sequence information on genetic resources (IDS). The framework is governed by Indigenous legal systems of Ecuador, UNDRIP, and Escazu Agreement, values, and anticipatory future systems thinking [10,11]. Kara-Kichwa Data Systems are shared here as a reference point for renascences – that is, place-based Indigenous re-emergences of lawful data systems – and dialogue, not as a universal template. The renaissances of Indigenous data authority would be sovereignty and self-determination embodied, embedded, enacted, and extended in LAC.

**Box 1. Key Terms:**

**Indigenous Anticipatory Futures Systems Thinking:** An intergenerational way of governing and decision-making that holds ancestral memory and future generations in active relationship. Futures-oriented systems governance beyond short-term optimization or prediction. It evaluates whether today's data use (e.g., AI training or genetic data [re]use) limits future Kara-Kichwa governance options or biocultural harms to future generations.

**Biocultural Engineering:** Collective, biocultural design of data, technologies, and institutions that sustain life across people, territory, and more-than-human relations. It is a practice-based technical design grounded in biocultural ethics, ecology, and collective benefit. It designs a community-controlled technology that supports sovereignty and rights of nature rather than extractive analytics.

**Indigenous Embassy:** A mandated community function for representing Indigenous authority (cultural leaders and translational leaders) in data actors' negotiations. It is a sovereign interface for cross-jurisdictional data diplomacy. They negotiate data equity shares agreements with universities, governments, or AI developers under explicit community mandate.

**Semipermeable Transboundary Membrane:** A governance boundary that allows selective, tier conditional data exchange. Controlled interoperability with external systems, allowing stewarded data access for research, development, and innovation (RDI) while prohibiting downstream reuse (e.g., AI training, consumer genomics, or commercial aggregation).

**Semantic Justice:** The right of Indigenous Peoples to define, name, interpret, and govern the meaning of their knowledge, information, and data. Authority over metadata, ontologies, and interpretation, preventing Indigenous concepts from being derogated, renamed, or reinterpreted in external databases without community consent. Guardian of data systems and data generation.

**Agentic Entity:** Recognition that communities, territories, and more-than-human relations exercise agency drawing from biocultural rights and ethics, not just individuals or institutions. Collective or non-human actors as rights-bearing entities, treating forests or waters as stakeholders in decisions about environmental data extraction and use.

**South Star in the Data Economy:** A re-oriented bio-digital and data economy guided by Indigenous law, reciprocity, and life regeneration rather than profit or revenue. A decolonial alternative to dominant data-capitalist value chains, reinstating Indigenous data authority for everything from holding to deriving value from Kara-Kichwa or Indigenous data, such as biodiversity or genetic data/DSI, back into community-led wealth, capacity, health, food, and ecological initiatives.

**Renaissances of Indigenous Data Authority:** Place-based re-emergences of Indigenous data systems grounded in each Indigenous Peoples' own law, institutions, and worldview. Plural, sovereign regeneration of data governance (not replication, universality, or diffusion). Other Indigenous Peoples, developing their own data systems frameworks, utilize it as a reference point or roadmap.

**Amařuk (serpent-dragon biokreature):** A fully elaborated Indigenous panarchic system – a multi-scale biocultural network, chimeric, biocultural data structure that encodes Andean-Amazonian memory, governs cross-scale information flows between Pacha (time/space) and Pacha Mama (territory), and the political and embodiment, embedment, enactment, and extending the adaptive cycle's capacity for both conservation of accumulated wisdom (information-based knowledge and wisdom-based knowledge (IBK+WBK)) and transformative renaissances/ renascences against systemic collapse.

## 1. Introduction

Indigenous Peoples and local communities (IPs&LCs) across Latin America and the Caribbean (LAC) *(Apŷa Yala)* are asserting their rights to access information and to participate in public decision-making, which are key state factors in any self-determination and sovereignty [12]. This article does not propose a universal model for all Indigenous Peoples or all contexts across LAC. Instead, it documents the Kara and Kichwa customary-law instrument designed to govern all data forms about/by/with/for Kara and Kichwa Peoples, in any format, medium, or expression. Hence, the Kara-Kichwa Data Sovereignty serves as a reference point for Indigenous data authority renaissances in LAC, which constitutes 45% of the Majority World, with low to no digital transformation [13]. "Kara-Kichwa" refers to the Kara Andean Nation and the Kichwa Andean-Amazonian Nation, Chinchay Zuyu (Ecuador). Each pillar embodies, embeds, enacts, and extends Kara-Kichwa legal systems, centering on relational logics, collective rights, and internal justice [26] enshrined in instruments such as the 2008 Ecuadorian Constitution [14], the UN Declaration on the Rights of Indigenous Peoples (UNDRIP), Escazu Agreement, and the rights of Indigenous Peoples to maintain and develop justice systems [11,15].

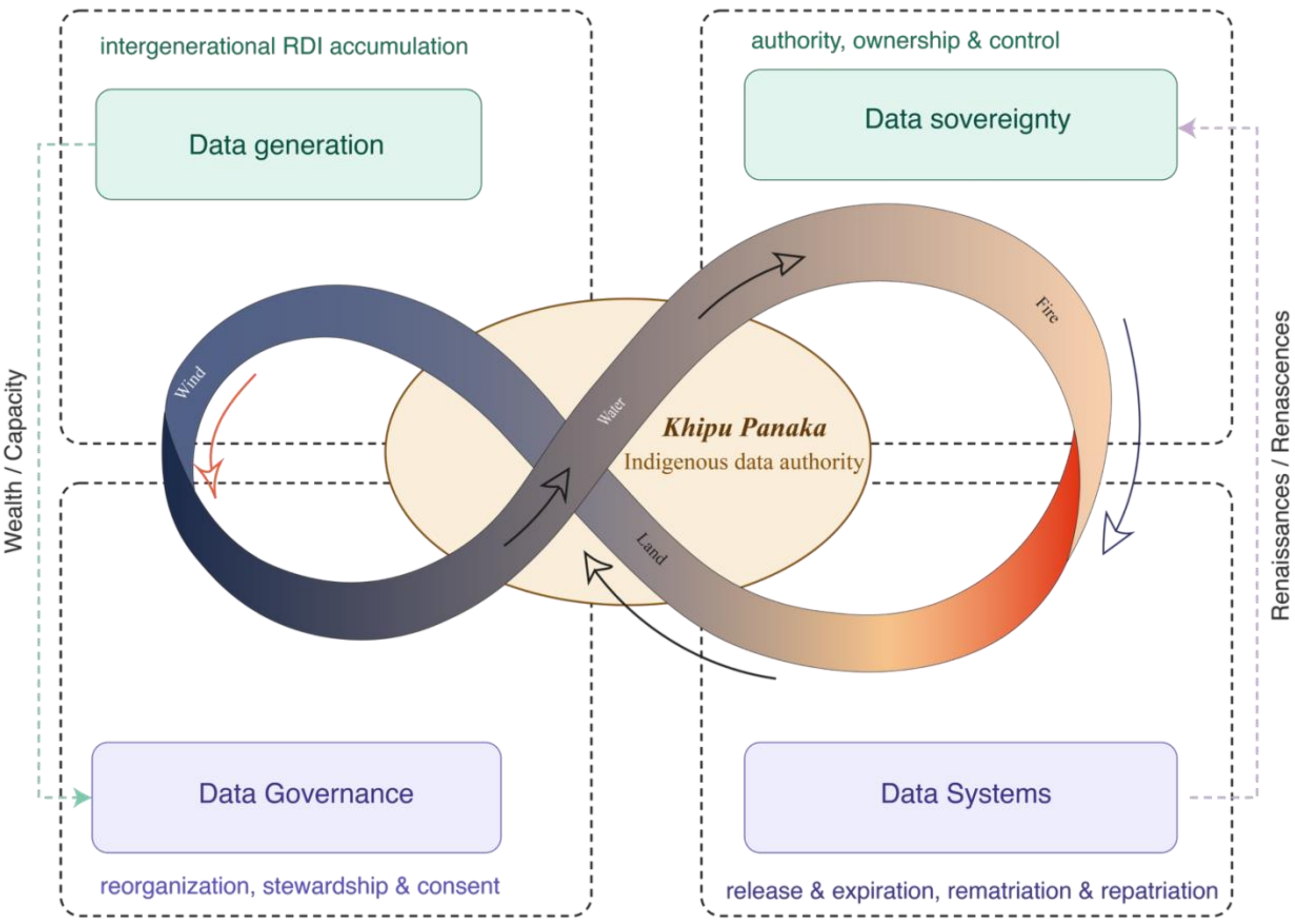

**Figure 1.** Kara-Kichwa Data Systems, inspired by *Amařuk* (serpent-dragon biokreature): Data type covered – **all formats, media, or expressions** about, by, with, for, originating, or generating from Kara-Kichwa, Indigenous Peoples, or local communities. Indigenous data, for the context of this framework, includes these classes – but is not limited to – digital, physical/paper, archival/legacy, bio-cultural/linguistic, biometric/identity, metadata/derived, AI/ML data, environmental, audio/visual/media, RDI data, or existing/emerging technology. Such data are not mere units in a database; they are extensions of **Khipu Panaka** – the living connecting thread across four phase nodes (nested levels of authority): (i) **Kara-Kichwa Data Systems** (the full scope of data originating from or concerning the commonwealth); (ii) **Kara-Kichwa Data Sovereignty** (the inherent rights and interests of Indigenous Peoples in the conceptualization, generation, collection, ownership, and application of their data, institutionalized nationally in the **2008 Ecuadorian Constitution**, regionally in the **Escazu Agreement**, and internationally **UNDRIP**); and (iii) **Kara-Kichwa Data Governance** (the principles, structures, laws, and practices through which the Indigenous commonwealth exercises control, ownership, and shareholdership over data across the full lifecycle – from generation (teal) to expiration (purple) – ensuring its management and use compliance with their rules, values, and collective interests for Sumak Kawsay, biocultural heritage, and community resilience). **Scope note**. Data systems in this framework are intentionally Kara-Kichwa to protect self-determination and semantic justice. **It is not a pan-Indigenous model and does not subsume other** Indigenous peoples or local community data regimes, which are governed by their own lawful authorities, epistemologies, and protocols. The specificity of this framework affirms, rather than excludes, the right of every People to govern their own data systems on their own terms.

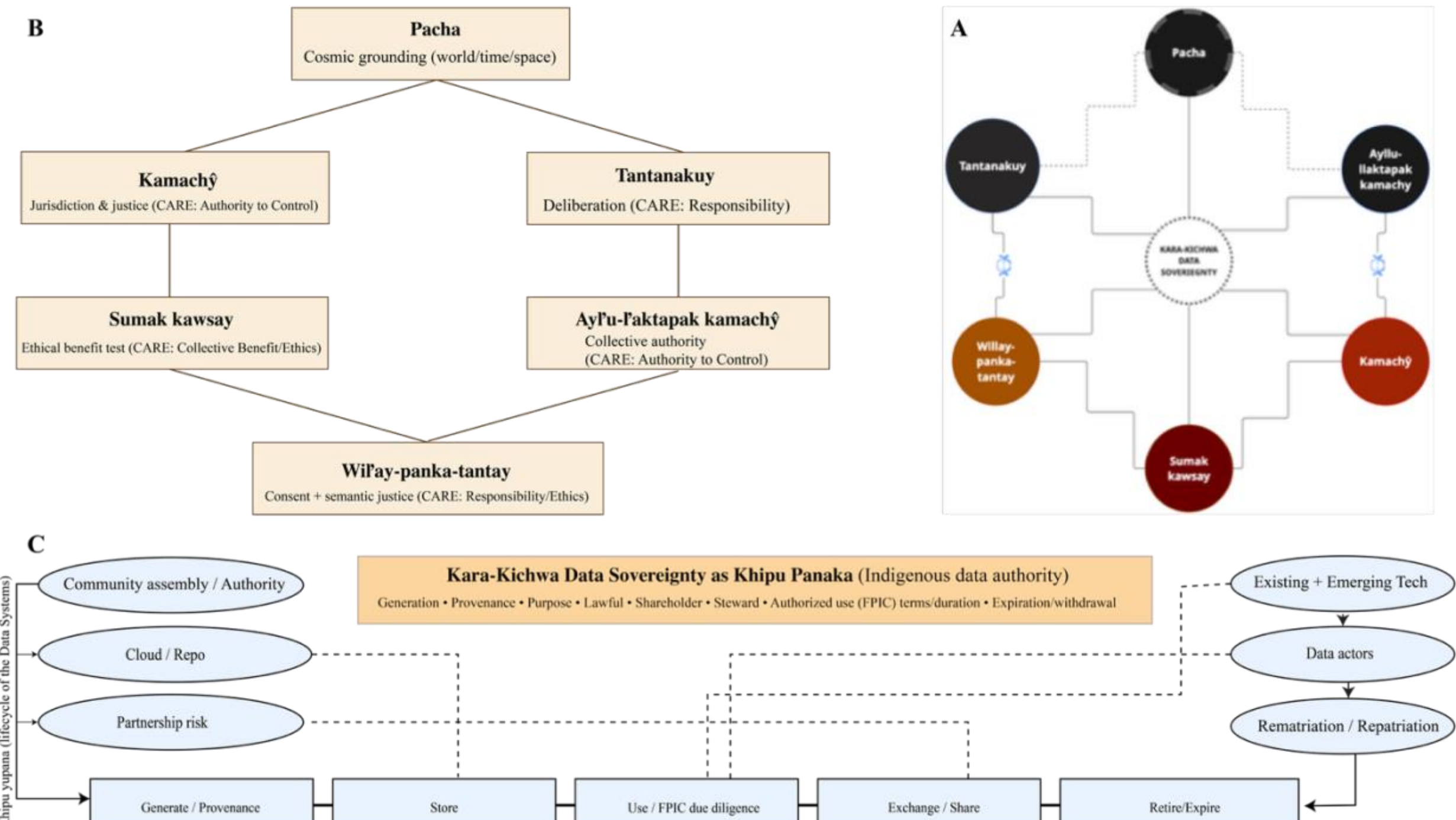

**Figure 2.** Kara-Kichwa Data Sovereignty Framework as Khipu Panaka Pillars. The framework is presented as a governance topology, not a linear workflow. It guides the lifecycle of data systems – **khipu yupana** – from generation to expiration (**panel C**). Nodes denote lawful governance functions grounded in Kara-Kichwa customary law (Justicia propia, **panel A**). Solid lines (–) mark binding governance process points; dotted lines (···) mark contingent pathways requiring heightened review (e.g., existing/emerging technology, data derivation, cloud repositories, partnerships/data actors). **Khipu Panaka** encodes provenance as relational logic metadata (provenance, purpose, lawful authority, steward, authorized use (FPIC) terms/duration, withdrawal/expiration) (**panel C**), **Pillars** aligning with CARE (Collective Benefit, Authority to Control, Responsibility, Ethics) (**panel B**), and supporting IEEE Std 2890 Indigenous provenance disclosure expectations. **Chakana**, sacred trans-systems geometry, connects **Khipu Panaka** interdependences across human, more-than-human, and institutional domains across time, space, and territory (**panel A**).

## 2.1. Framework Development Approach

The Kara-Kichwa Data Sovereignty Framework emerged from the authors' lived governance experiences and perspectives on Kara-Kichwa or Indigenous communities as they navigate RDI amid digital transformation and emerging data economies. Its development is grounded in what Wilson (2008) calls relational accountability: Indigenous research methodology is not a set of external tools applied to a community but an autopoietic practice of relationship, obligation, and accountability to all participants in a knowledge system [16]. It emerged from the BioKulture Systems Design (BKSD) praxes of the right to story and place, as well as Indigenous storytelling, not from academic exercise [17,18]. The governance architecture reflects the deliberative practices codified in the Kara-Kichwa justice systems [19] and the rhetoric of Indigenous institution theory [20], which are articulated in the Ecuadorian Constitution, UNDRIP, the Escazu Agreement, and GIDA. Autopoietic, relationship-based dialogues with knowledge keepers, cultural and translational leaders, elders, and community-elected authorities form the pillars that govern Kara-Kichwa, or Indigenous data, reflecting what Kovach (2021) calls conversational methodology [21], rather than helicopter research protocols [22]. It enacts intergenerational governance in the authorship itself, including a knowledge keeper, an elder, a cultural leader, Mama María Rosa Panama, and a representative for future generations, KunTikzi Flores.

To strengthen external interoperability without diluting Kara-Kichwa or Indigenous authority, we conducted an Indigenous-informed structured review mapping between the pillars and a few selected Indigenous and international governance instrument examples (e.g., GIDA-CARE, OCAP™, Māori Data Sovereignty, Africa Data Ethics, UNESCO AI ethics) [23], prioritizing frameworks that: (1) are Indigenous-led or Indigenous-accountable, (2) provide operational mechanisms for consent, provenance, and authority-to-control, and (3) address contemporary risks in AI, digital repositories, and genetic/Digital Sequence Information (DSI) governance. This process produced a bidirectional alignment: Kara-Kichwa law remains the decision-making authority, with external sources serving as reference points for negotiating ethical, legal, and technical interoperability with data actors (academia, industry, public and private sectors, etc.).

## 2. Kara-Kichwa Data Sovereignty Pillars

The Kara-Kichwa Data Sovereignty Framework builds on five core pillars of customary legal language, each embodied, embedded, enacted, and extended in key aspects of Indigenous data governance. They aim to maintain the relational, genealogical, and kinship obligations of the data [re]generation throughout its journey across the data systems (Fig. 1), addressing the full spectrum of data flows and intergenerational responsibilities (Fig. 2). They function as relational logics (*Chumpŷ*) that enact *Khipu Panak*a to secure data sovereignty, auditability, ownership, self-determination, and the self-actualization of *Sumak Kawsay* for the Community, *Rhunā-kuna* (living systems), and *Pacha Mama* (Rights of Nature) libraries (wealth/capacity) – we inhabit (Fig. 1). These living pillars serve as renaissances/renascences reference point for Indigenous Data Authority in Latin America and the Caribbean.

To support practical uptake by community partners, governments, repositories, and researchers, we operationalize the pillars in the data governance architecture across the data systems (Fig. 3): (1) proposal and consent, (2) collection and documentation, (3) storage and stewardship, (4) access and use, (5) shareholders (equity-sharing) and accountability, and (6) expiration/retirement and memory care (rematriation/repatriation). Decision authority is exercised through community assemblies (Tantanakuy) – for Kara-Kichwa data, it is the Kara & Kichwa Peoples, and for non-Kara-Kichwa data, it is Indigenous Peoples from whom the data originates – designated data stewards (Wilʼay-panka-tantay), and community-elected authorities who carry jurisdictional responsibility (Aylʼu-lʼaktapak kamachŷ). Negotiations occur only through clearly mandated representation (e.g., data curation council and/or an Indigenous embassy function) (Kamachŷ) to prevent individualization of consent, to safeguard community assembly remains the governing authority, and to ensure that relational obligations – human and more-than-human – remain binding (Sumak kawsay).

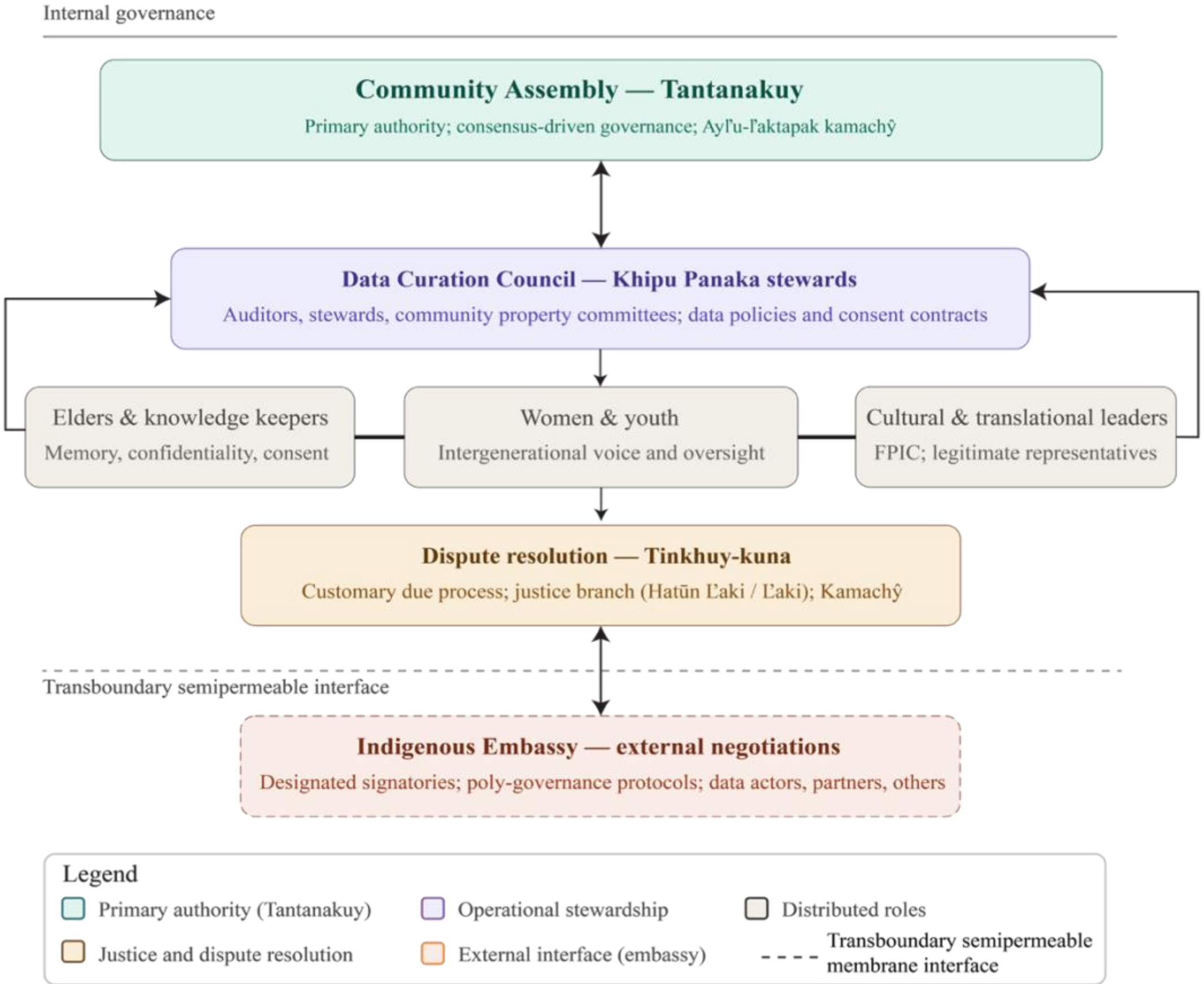

Figure 3. Kara-Kichwa Data Governance Architecture. The primary decision-making body is the **Community Assembly (Tantanakuy)**, which exercises **collective authority (Ayl'u-l'aktapak kamachŷ)** over all data matters through consensus-driven deliberation, in accordance with the Ecuadorian Constitution (Art. 66(19), 57, 71-74, 92, and 171), UNDRIP (Art. 3-5, 15, 18-19, 31-33), and Escazu Agreement (Art. 6-9). The **Data Curation Council** holds operational authority over data policies, access consent contracts, and the rule of law for data retraction and use (Pillar 1.2). This council consists of elders and knowledge keepers, women and youth exercise intergenerational oversight and representation; and cultural and translational leaders, holding FPIC signatory authority in engagements with data actors **(Wil'ay-panka-tantay)** (Pillar 3.1; UNDRIP Art. 19; Ecuador Constitution Art. 57; Escazu Agreement Art. 7). **Disputes and data governance offenses (Hat'ūn Ľakī or l'akī)** resolve through **Tinkhuy-kuna**, a customary procedural law and the Indigenous Justice Branch of Government

## 0.0 Kamachŷ | Self-determination and Authority

**0.1 Control**. Indigenous Peoples have the sovereign authority to govern the relationship to all data types about/from/by/for Indigenous communities (Fig. 1). This encompasses the inherent right to own, audit, benefit from, and possess data about themselves, their communities, and their resources. Control extends to disputes and data offenses in conformity with l'akī (*llaki*), and to determining how such data is generated, collected, analyzed, shared, and retired. It affirms sovereign exercise of decision-making power as recognized in UNDRIP Art. 31; Art. 57 in the Ecuador Constitution; Art. 7 in the Escazu Agreement, which provides Indigenous Peoples' right to "maintain, control, protect and develop" their biocultural networks (e.g., biocultural assets) [24].

**0.2 Jurisdiction**. Indigenous Peoples have sovereignty over data rulemaking (jurisgenesis), resolution of offenses (l'akī) via community procedural due process & due diligence [19], and Hat'ūn l'akī enacts outward jurisdiction coverage on disputes and data offenses, asserting collective authority, auditability, and accountability standards in all forms of data and in any format, medium, or repression (Fig. 1) (UNDRIP Art. 34; Escazu Agreement Art. 8; Ecuador Constitution Art. 171) [25]. Data stored under Indigenous authority ensures that customary laws and values govern data handling (Khipu Panaka stewards: consent and refusal. Ex-situ data repositories shall be governed by Indigenous embassy negotiations in data provenance disclosure expectations (IEEE std 2890), and equity shares (Fig. 3). The repository/cloud follows TRUST principles (Transparent, Responsible, User-focused, Sustainable, and Technology), sustaining community oversight [26].

**0.3 Autonomy**. Indigenous Peoples operationalize the political autonomy recognized in nation-state treaties (e.g., Ecuador's constitution), international UNDRIP agreements, and regional Escazu Agreement in any initiative affecting their biocultural assets: land, data, and resources, in any format, medium, or expression. The FPIC political participation mechanism has become an increasingly internationally recognized instrument for community-government-society-private partnership, extended to data governance in the data economy, and a key component of self-governance [27]. Autonomy enacts its right to use data for its own governance, planning, and decision-making. Data categories shall serve their own RDI and community priorities (Fig. 2).

### 1.0 Ayl'u-l'aktapak kamachŷ | Collective Authority & Polygovernance

**1.1 Commoning**. Collective action and communal governance focus on commoning data equity shares based on intergenerational responsibilities. It develops Indigenous institutional systems, facilities, homes, and economic activities. Collective rights govern data through embodied, embedded, enacted, and extended participation and management of community members working together to uphold the dignity and unity of the commons (Ecuador Constitution Art. 57; Escazú Agreement Art. 7). It brings Peoples together to engage in biocultural engineering, to develop and maintain their data infrastructure and governance processes, and to define Indigenous websites and media, UNDRIP Art. 16 [15].

**1.2 Transfer**. Data is not left to "experts" alone. Instead, khipu panaka stewards: knowledge keepers, elders, youth, women, and community-elected authorities all have roles in governing data. It operationalizes the transfer of power (i.e., community data curation councils, data auditors, data stewards, community property committees) and tasks them with creating data policies (protocols for RDI, data access consent contracts, and the rule of law for retraction, etc.). Indigenous embassies negotiate with data actors to build their workforce and safeguard the generation, collection, processing, and analysis of data for governance, equity shares, and ownership or co-ownership (Escazu Agreement Art. 6; UNDRIP Art. 17; Ecuador Constitution Art. 1) [28,29].

**1.3 Connectivity**. Data flows through trans-systems of a semipermeable transboundary membrane enacting sovereign external participation (Fig. 3). In engagements with other Indigenous communities and data actors, Indigenous Peoples maintain an equal partnership and a clear Indigenous embassy within polygovernance protocols [29,30]. The connection between Indigenous Peoples, both internationally and nationally, catalyzes strategies and infrastructure to strengthen these living pillars, serving as renaissances/renascences reference point for Indigenous Data Authority via trade routes, facilitating resource exchange, advancing commoning goals, and institutionalizing IEEE Std 2890, data auditors, data stewards, and community property [31].

## 2.0 Tantanakuy | Collective Deliberation & Relational Accountability

**2.1 Interdependence**. In the ethos of Tantanakuy, Tinkhuy-kuna is a collective dialogue, deliberation, and responsibility-setting process, grounded in wil'aychina (right to story and place), tapuychina (generate RDI), al'ichina (reconciliate), and katichina (mentorship) with 'love,' 'care,' and 'tenderness' to reinstate Community (human and nonhuman) harmony [19]. This interdependence is guided by biocultural ethics among peoples, cultures, and nature, embedding the functionality and operationality of data systems. The embodiment of interdependence may articulate the relational and genealogical nature of data, thereby generating biocultural relationality (biocultural assets) articulated by Art. 171 and Art. 71-74 of the Ecuadorian Constitution [14].

**2.2. Relational Accountability**. Any data involving Indigenous Peoples and their commonwealth shall be handled with dignity, in accordance with intergenerational obligations and FPIC due diligence. It guides data actors away from nearly obsolete "good-faith pledges" of compliance toward developing understandable hybrid agreements (customary law + contractual elements, licenses, etc.). Inherently, it builds on the paktachina (execution of collective resolutions) that demands patience and diplomatic dialogue: the community's 'tenderness' (tending and tension) shall align with the proposed project(s); otherwise, "bad faith" data actors can jeopardize relational accountability in the collective deliberation of data governance and shareholdership. It may result in FPIC due diligence violations, with data ramifications and implications [27].

**2.3 Reciprocal Responsibility**. Khipu Panaka stewards, or Community guardians, ensure Indigenous data and big data (the operating system of the territory) serve the collective benefit through the production of place-specific understanding and the collective processes. It maintains 'us-all' in data kinship, reminding everyone that behind every data point, data collection instrument (data actors), and method is in relational logics, and 'do no bio-cultural harm' is the basis for avoiding data that stigmatizes or blames and practices collective rights and privacy (Pillar 0.0). Data actors should implement IEEE Std 2890 to avoid data dispossession justifications under the doctrine of discovery and cherry-picking findings that distort the ground truth and undermine social validation (Rhunā shimī) [32].

## 3.0 Wil'ay-panka-tantay | Knowledge Confidentiality & Ancestral Memory

**3.1. Consent**. The Community holds ultimate decision-making power with the legitimate Indigenous governance body in their language, time, space, and land. It exercises this power in the due process for Free, Prior, and Informed Consent and Good Faith (FPIC+GF), and due diligence regarding the political autonomy of Indigenous Peoples, across all data forms and activities throughout the lifecycle of data systems (Fig. 2) [27]. As mandated by the Sarayaku v. Ecuador ruling, the PFIC+GF process shall be a genuine dialogue of mutual trust, aiming to reach an agreement based on truthfulness, transparency, the reality of conditions, and free of deceit, allowing data actors to obtain explicit consent from the community assembly led by Khipu Panaka stewards, UNDRIP Art. 19, Escazu Agreement Art. 7, and Ecuador Constitution Art. 57 [14,15].

**3.2. Semantic Justice**. The Community guards translations and interpretations of Indigenous data, covering all system forms in any format, medium, or expression. Indigenous data reflect their languages, terminologies, worldviews, values, and social realities (UNDRIP Art. 11; Escazu Agreement Art. 6). Indigenous Peoples govern databases, datasets, and archives, and the use of such data requires community permission, restrictions, and oversight (Ecuador Constitution Art.

57(12-13)). Indigenous Peoples have the right to maintain, control, protect, and develop their cultural knowledge and intellectual property (UNDRIP Art. 31).

**3.3. Authenticity**. The Community makes all the calls: expectations for Indigenous data provenance disclosure (IEEE std 2890), authenticity (the resistance & survivance of characteristics, the quality of tangible heritage, and the understanding & honoring of intangible traditions and practices), and decision-making power rests with the Community (for Kara-Kichwa data, it is Kara-Kichwa. For non-Kara-Kichwa data, it is Indigenous Peoples from whom data originates) [18,33]. Indigenous embassy (designated signatories) grounds context of authenticity, guiding authority, physical origin, legitimacy, free of conflict of interest, extent of negotiation reach (Ecuador Constitution Art. 57(13); UNDRIP Art. 32; Escazu Agreement Art. 9). It counteracts Indigenous pretenders, individual consent, etc.

**4.0 Sumak kawsay | Biocultural Ethics & Intergenerational Responsibility**

**4.1. Guardianship**. It prevents tokenism or symbolic engagement with Indigenous data obligations. Data actors shall demonstrate the real influence of Khipu Panaka stewards or Community Guardians over everything in the lifecycle of Indigenous data systems (Fig. 2). Indigenous embassy guides the design of diplomatic partnerships, and ongoing Indigenous polygovernance is non-negotiable for cloud or local repositories transparency, intergenerational responsibility, Indigenous user focus, sustainable resource use, and just technology. It reaffirms advances in dignity-centered development, institutional implementation (human & nature rights), systems evolution, and the resilience of communities and their environments (UNDRIP Art. 20; Escazu Agreement Art. 1; Ecuador Constitution Art. 51).

**4.2. Biocultural ethics**. Data should advance Sumak Kawsay, biocultural rights articulated for collective well-being, equity, and dignity, affirmed or recognized in the Rights of Nature (RoN) (Pacha Mama) Ecuador Constitution Art. 71-74, 32, 322; Escazú Agreement Art. 8; UNDRIP Art. 11, 13, 25, 26. Biocultural ethics closes the authority enforcement gaps of all data forms for shareholdership (equity shares), authority, ownership, and use [34], reframing the paradox of ethics "'Golden Rule', to treat others as you wish to be treated, and the 'Platinum Rule' of treating others as they wish to be treated" [35].

**4.3. Worldviews**. The knowledge (IBK) and 'understanding' production (WBK) on dimensions of relationality and genealogy of data from Nature, Culture, and Institutionality rest with the authority power of the Community, extending to Indigenous data in internal or external onboarded data in existing and emerging technology (UNDRIP Articles 1-5, 13, 34; Escazu Agreement Art. 7). Kara-Kicha and non-Kara-Kichwa worldviews may be articulated in biocultural rights and ethics, reaffirmed in RoN (more-than-human) and community (human) are rights-bearing [36]. The authority on data-derived value, not limited to ownership, lies with the community that maintains and regenerates life cycles, not with those who exploit or commodify natural assets.

## 3. Alignment with External Frameworks and Standards

As noted in subsection 2.1, these few example resources discussed below provide complementary mechanisms and references not as validators over Kara-Kichwa law, but as translation interfaces to support interoperability, negotiation, and accountability across institutions.

**IEEE Std 2890-2025** establishes a common set of parameters for data provenance of Indigenous Peoples and their territories, ensuring the implementation of **GIDA-CARE Principles** (Collective Benefit, Authority to Control, Responsibility, Ethics) [37] and the compliance disclosure of their relationality and genealogy in all data (human and more-than-human) [38] by the multiple data actors. Kara-Kichwa, or Indigenous data authority embedded in the pillars, guides data actors (e.g., repositories or institutions) in storing Kara-Kichwa or Indigenous data (e.g., using existing or emerging technologies) that to conform with Indigenous provenance standard, the dataset shall carry relationality and genealogy disclosures: who the lawful authority is (community assembly/designated stewards), the consent conditions and duration, the data's origin and purpose, restrictions on downstream use (AI training, consumer genomics, [re]use), and mechanisms for withdrawal/expiration. In Kara-Kichwa terms, this is the Khipu Panaka: a provenance braid that binds technical metadata to lawful accountability.

**OCAP™ (First Nations: Ownership, Control, Access, Possession)** principles serve as a national foundation for Indigenous data sovereignty. The Kara-Kichwa pillars align with Ownership and Control, and address controlled Access and exchanges, emphasizing community Possession of data infrastructure. Andean-Amazonian communities shall govern databases, surveys, and archives about them and their commonwealth, thereby guaranteeing "collective intellectual property of their ancestral knowledge," as stated in the Ecuadorian Constitution, Art. 57, 92; UNDRIP, Art. 31; and the Escazu Agreement, Art. 7 [14].

**Māori Data Sovereignty** principles have been linked to the Māori duties of *Taonga* (treasure) [39], which are shared with the Kara-Kichwa *Kurhī* (sacred/treasure) due diligence systems, embodied, embedded, enacted, and extended across the 5 pillars (Fig. 2). Likewise, these pillars contribution aligns with the **African Data Ethics** outline principles of "Assert Data Self-Determination," "Utilize Communalist Practices," and "Invest in Local Data Infrastructures" [40]. Contributions to the Majority World may help close the ~45% digital transformation gap, thereby improving the negotiating conditions for IPs and LCs [7,13].

**The 2021 UNESCO** Recommendation on the Ethics of Artificial Intelligence (AI) (adopted by 193 states) includes autonomous participation and data governance by Indigenous Peoples in AI development [41]. Kara-Kichwa or Indigenous data authority safeguards the development of existing and emerging technologies with Indigenous Peoples, rather than imposing them upon them (see Big AG & Big Tech) [42]. The UN Permanent Forum on Indigenous Issues (UNPFII) affirmed IPs' role in AI governance at its 24$^{th}$ session. In mountain societies like Ecuador, the lack of an Indigenous data sovereignty policy has created structural barriers to operationalizing Free, Prior, and Informed Consent (FPIC) due process in the National Statistical Commission for Indigenous, Afro-Ecuadorian, and Montubio Peoples [24], excluding IPs&LCs from decision-making in the sustainable mountain development [7].

The establishment of a subsidiary body under Article 8(j) and other provisions of the Convention on Biological Diversity (CBD) related to IPs&LCs introduces modalities that should guide due diligence for the pillars of Indigenous data sovereignty in all data forms in any format, medium, or expression for everything from holding to deriving value from Kara-Kichwa or Indigenous data [24,43,44]. This framework seeks to reinstate Indigenous data authority to improve the negotiating conditions for securing benefit-sharing CBD mechanisms alongside Local Context labels (TK) and notices (BK) recommended in DSI monitoring in the Cali Fund [45,46]. The pillars offer the architecture blueprint for data authority in contracts, licenses, and treaties with data actors.

## 4. Indigenous Data Authority in Collaborations

The Kara-Kichwa Data Sovereignty Framework articulates the obligations affirmed or recognized in UNDRIP, the Escazú Agreement, the Constitution of Ecuador, or the Global Indigenous Data Alliance (GIDA) recommendations for everything from holding to deriving value from Kara-Kichwa or Indigenous data. Data actor allies (e.g., academic, industry, public) may collaborate when partnerships are structured as co-creation under Kara-Kichwa or Indigenous authority. Collaborations shall specify: (1) purpose limitation (what questions the data may answer), (2) consent continuity (how consent is renewed, revised, or withdrawn), (3) tiered access (what remains confidential, community-only, or negotiated for controlled external access), (4) provenance and attribution (Kara-Kichwa or Indigenous semantic justice in interpretation), (5) benefit-sharing and reciprocity (capacity fortification, infrastructure, resources, authorship norms, and deriving benefits), (6) use restrictions (explicit permission requirements for AI training, model release, and secondary use), and (7) disputes and data offense (where bio-cultural harms occur – misinterpretation, stigmatization, unauthorized [re]use – Kara-Kichwa or Indigenous authority retain the right to retract consent, require remediation, and pursue accountability through customary and applicable hybrid legal agreements).

## 4. Conclusion

The Kara-Kichwa Data Sovereignty Framework represents a synthesis of semantic justice within Kara-Kichwa customary law and justice systems, articulated in ways that can interface with international, regional, and national governance instruments. The five living pillars serve as renaissances/renascences roadmap for the Indigenous Data Authority for Kara-Kichwa and other Indigenous nations who choose to reference on their own lawful terms to govern the lifecycle of data systems in service of collective well-being and social–ecological futures.

For readers unfamiliar with them, the Sustainable Mountain Development agenda and the Science Panel for the Amazon are cited here because they identify major data gaps and governance needs in the Pan-Andean–Amazonian-Atlantic region; our intervention is that closing those gaps without Indigenous data authority reproduces extraction. In Kara-Kichwa governance, data is not merely a commodity: it is collective identity, territory memory, kinship network, and sacred/treasure *(Kurhī) – a living, breathing, intergenerational sacred trust*. Like data that holds relationality and genealogy to the physical space, the pillars embody, embed, enact, and extend within Pacha Mama, reminding 'us-all' of relational accountability to the living world – Pacha Mama as a rights-bearing, life-generating commons – not an abstract metaphor. Therefore, any data use that undermines community authority, dignity, ecological regeneration, or RoN obligations violates the lawful purpose of data stewardship and triggers governance response, including consent retraction and remediation.

The pillars support efforts to advance dignity-center development, closing enforcement gaps through RoN resolutions in conservation [34]. Where the UN Sustainable Development Goals (SDGs) emphasize "leaving no one behind," Kara-Kichwa Data Sovereignty asserts that true prosperity is achieved only when 'us-all' enjoy development and share the duty of care to sustain life [7]. It articulates FPIC due process and IEEE Std disclosure practices in RDI [27], synthetic biology, and AI [47,48], serving as the South Star for data-economy mechanisms on the Andean-Amazonian-Atlantic governance affecting +30 million mountain residents [8].

The journey ahead requires continued vigilance to ensure that, whether dealing with RDI, a government statistician, policymaker, or an AI algorithm processing Indigenous data, Indigenous Peoples can stand firm on their terms: *Our data, like our lives, will be governed by our laws, values, and our vision of Sumak Kawsay and Rhunā-kuna*. This is Indigenous data authority in action, a powerful affirmation that, in the age of digital transformation, Indigenous Peoples will not be reduced to mere data aggregates or AI inferences but will be authorities, authors, and owners of their sovereign narratives, from time immemorial, now, and into intergenerational futures.

This framework provides a language architecture for cultural and translational leaders and the Indigenous embassy, leads the next era of agrifood systems, empowers women's associations to secure funding for maternal health clinics, and implements Indigenous data authority to close data policy gaps in existing and emerging technologies. The authenticity of *'nothing about us without us'* is *kreolized* among Khipu Panaka stewards, who govern Kara-Kichwa or Indigenous data; therefore, Indigenous Peoples govern their sacred/treasure (data) biocultural heritage for the present, intergenerational future, and past. This renaissances/renascences roadmap may support the fulfillment of self-determination, development, and the survivance vitality of Indigenous cultures and laws in the age of digital encounter.

**Funding**: The Authors report no funding for this research.

**Data Statement**: No Data were produced in the review article besides the data cited throughout the in-text citations. No data are reported or made available, and all referenced data are cited in the paper.

**Acknowledgments**: The authors are grateful to the Indigenous and settler scholars who provided the reviews cited in this article. Additionally, we acknowledge the contributions of the libraries that are the territories we inhabit.

**Conflicts of Interest**: The author declares no conflict of interest.

**Author(s)**: The authors contributed equally to this paper. We apply Indigenous anticipatory futures systems thinking and the United Nations REV3 Declaration on Future Generations to include KunTikzi Flores as a co-author of this paper, who represents future generations and the intergenerational responsibility of Indigenous Peoples in LAC. In the Indigenous commonwealth rebuilding process, we return to our power source (i.e., culture, language, spirituality, and philosophy) inside the Kara-Kichwa *Pacha* to reenergize and develop new approaches to address new and unfinished challenges; thereby, we include Maria Rosa Panama as co-author of this paper for gatekeeping our natural laws and all the constitute our culture, ancestry, and rights, who represents the authorship of nature and the Indigenous Republic.

Using the Inka history in knots, the authors provide counter-mythology to technology for data management and governance. This book, in recent times, provide examples of the Andean technology sophistication to store, manage, and interpret data before WEIRD computers.

Using the Khipus khipu boards, and sacred texts, the authors articulate Khipu Panaka (Indigenous data authority) and Kurhī (data as sacred/treasure) of the data intergenerational responsibility along the relational and genealogical data in Andean-Amazonian worldviews. It provided recent scholarship on the sophistication, legitimate, value, and ontology of technology, data systems, and information.

Using Mundos de creacion, the authors link how different interpretation of natural law (jurisgenesis) in the language of Indigenous Peoples encoded in data actors. The book provides clear examples of data actors (devices) used to encode Indigenous data across Latin America.

Using the remapping science, authors argue the persisting WEIRD colonial legacies in science and its productions and how it calls for novel science diplomacy led by communities. It is one of the most recent studies linking history and its ramification and implication in the Majority World science.

Using the CONAIE-IPRI, authors make the case for the relevance of the Kara-Kichwa Data Sovereignty. It provided ground-truthing testimonies on how the lack of Indigenous data authority and governance disproportionally ramifies and implicates vulnerable Indigenous members.

Using the UN sustainable mountain development, authors connect the lack of Indigenous data governance on Andean data for their own governance, making the development plans external data actors' dependencies. The report is one of the most updated UN assessments of Andean societies.

Using chapter 24, authors argue how WEIRD science may have isolated Andean communities from political participation on the Andean-Amazonian-Atlantic governance. The book assesses the current climate, biodiversity, and social treats to the region.

Using unlearning modernity, authors introduce to the text the arguments of WEIRD, Majority World, and Minority World to counter-mythology of western science. It assess how IPCC might have fail to handle respectfully Indigenous and Local Knowledge.

Using framings, authors ground the inherent nature of Indigenous anticipatory futures systems thinking that grounds intergenerational responsibility of Indigenous Peoples on their practices and worldviews. It is the study with multiple Indigenous participates looking into their multi-generational life plan management.

Using the UN rights of Indigenous Peoples, authors root this framework in international agreements on the distinctive recognition of Indigenous Peoples rights to exercise their own legal systems. This report reaffirms international UNDRIP for its exercise within communities and external actors.

Using the Escazu Agreement, authors argue of the existing and current regional agreements on the assertion of data governance by Indigenous Peoples and local communities in Latin America and the Caribbean. It provides regional obligations context in multi-lateral agreements to manage participation, justice, and access of the communities in information pertaining their interests.

Using the digital transformation, authors links how the majority world, particularly Latin America, remains with low or no digital transformation in the 21st century. It is a comprehensive study on the directions of majority world in the digital (data) economy.

Using the constitution of the Republic of Ecuador, authors reaffirm in the framework the national recognition of Indigenous systems or institutions, grounding obligations within this framework in the Ecuadorian territory. The 2008 constitution, historically, recognize the inherent rights, legal systems, authority, and jurisdiction of Indigenous Peoples.

Using the UNDRIP, authors strengthen the international agreements on the inherent rights of Indigenous Peoples data authority to own, possess, control, govern, retract, refuse, and consent. It is an international instrument for rights-based community-government-society-private partnership.

Using Research is ceremony, authors grounds an Indigenous scientific methodology on the development approach of this framework. It is an important articulation of Indigenous research methodology in a non-fetishization of decolonial or participatory WEIRD methods.

Using BioKulture Systems Design, authors strengthen the Indigenous scientific methodology giving a Kara-Kichwa context to their inherent right to story and place. It is one of the few Indigenous scholarships that elucidates a tripartite RDI design partnership.

Using Indigenous storytelling as research, authors link the Indigenous scientific methodology implemented in this framework to other areas of research through the legitimacy of storytelling. It is a vital work to elucidate the legitimacy of storytelling as a medium or expression for imperial data.

Using UNORCAC, authors ground the pillars of the framework in the Indigenous Ecuadorian legal systems and internal justice (justicia propia) of the Kara-Kichwa or Indigenous Peoples of the Ecuador. It is a first-of-its-kind Indigenous justice manual developed the Indigenous communities of Cotacachi that complements internal exercise of justice with the constitution of Ecuador.

Using Indigenous institutional theory, authors link the concept of Indigenous rhetorics to the Indigenous scientific methodology, centering a level of institutionality to the practice of autopoietic dialogue or conversations. It is a vital scholarship contribution to the theorization of what is Indigenous institutions.

Using Indigenous methodologies, authors fortify the Indigenous scientific methodology applied to this framework. The element of conversation method is a significant contribution of this scholarship to Indigenous research methodology and storytelling for empirical data.

Using ending domestic helicopter research, authors connect to historical academic exercise in the development of perhaps similar framework. It is an significant contribution to how scientific knowledge has been generated.

Using Indigenous-informed scoping review, authors support the Indigenous scientific methodology implemented by literature review of selected few examples on scholarship that support Indigenous data sovereignty, authority, and self-determination. It is an vital for how literature review for proceeds from the perspectives of Indigenous Peoples.

Using EMRIP, authors present a case of the lack of Indigenous data governance in the Ecuador national statistics. It is a foundational work in Indigenous data globally, highlighting the gaps in data collections, aggregation, and disaggregation.

Using the Navajo courts, authors argue the inherent autonomy of Indigenous community in jurisgenesis, particularly, in data rulemaking. It presents multiple analysis of the formalization of Indigenous Peoples law making.

Using TRUST, authors emphasize the technological aspect of data stewardship to build trust and maintain community property. It focuses on technological sustainability and the user of data.

Using IUCN, author link the framework to emerging instruments of biocultural ethics. The motion request the operationalization of the rights of nature instrument in the communities.

Using Indigenous justice, author further connects the framework to operationalize biocultural ethics within the framework. It is a significant contribution on the counter-mythology to WEIRD ethics.

Using biocultural rights, authors elaborate the biocultural ethics sub-pillar according to multiple territory case studies, leading the pathway to application of the biocultural rights, assets, and sovereignty. It is a foundational book on biocultural dimensions of Indigenous Peoples.

Using operationalizing CARE and FAIR, authors make the case for the relevance of the framework with internation development of Indigenous data sovereignty principles by GIDA. This study links the clear orientations of CARE to complement the default "open" data of FAIR.

Using the IEEE, authors link the framework to the common parameters to disclose (meta)data provenance to the inherent data systems of the Kara-Kichwa or Indigenous Peoples. This is the first of its kind standard developed by Indigenous experts to operationalize CARE principles across data actors.

Using the Māori data sovereignty, authors center the duties generated by the Indigenous Peoples to their governance data systems. It provides Indigenous Nation level data sovereignty layers.

Using Africa data ethics, authors argue to how this framework can guide the digital transformation of the majority world if are to enfranchise in the data/digital economy. It is a foundational data ethics contribution to the majority world.